\documentclass[conference]{IEEEtran}
\IEEEoverridecommandlockouts
\usepackage{graphicx}
\usepackage{colortbl}
\usepackage{float}
\usepackage{amsmath}
\usepackage{enumerate}
\usepackage{notoccite}
\usepackage{authblk}
\usepackage{color}
\usepackage{tabularx}
\usepackage{gensymb}
\usepackage{listings}
\def\BibTeX{{\rm B\kern-.05em{\sc i\kern-.025em b}\kern-.08em
    T\kern-.1667em\lower.7ex\hbox{E}\kern-.125emX}}

\IEEEoverridecommandlockouts\IEEEpubid{\makebox[\columnwidth]{978-3-903176-32-4~\copyright~2021 IFIP \hfill} \hspace{\columnsep}\makebox[\columnwidth]{ }}
\begin{document}
\title{
	Analysing Design Approaches for the Power Consumption in Cyber-Physical Systems}

\author[1]{Patrizia Sailer}
\author[1, 2]{Igor Ivkic}
\author[1]{Markus Tauber}
\author[3]{Andreas Mauthe}
\author[2]{Antonios Gouglidis}
\affil[1]{{\normalsize \textit{University of Applied Sciences Burgenland} - Eisenstadt, Austria}}
\affil[2]{{\normalsize \textit{Lancaster University} - Lancaster{}, United Kingdom}}
\affil[3]{{\normalsize \textit{University of Koblenz-Landau} - Koblenz{}, Germany}}
\renewcommand\Authands{,  }

\maketitle

\begin{abstract}
	The importance of Cyber Physical Systems (CPS) and Internet of Things (IoT) applications is constantly increasing, especially in the context of Industry 4.0. Architectural decisions are crucial not just for performance, security and resilience reasons but also regarding costs and resource usage. In this paper we analyse two of the fundamental approaches to design control loops (i.e. time-driven and event-driven), show how they can be realised and evaluate their power requirements. Through this the design criteria can be extended also considering the optimization of energy related aspects.
\end{abstract}

\section{Introduction}

In recent years, advances in computing technologies have had a profound influence in the development of the fourth industrial revolution \cite{ref01}. The so-called Industry 4.0 is driven by Cyber-Physical Systems (CPS) and the Internet of Things (IoT), which are tightly integrated in and interacting with the physical world \cite{ref02} \cite{ref03}. A CPS is formed by components that are capable of communicating with each other, measuring their environment, analysing those measurements and set actions to change environmental conditions. The IoT could be considered as the backbone of a CPS, which is responsible for connecting IoT devices, sensors and actuators~\cite{ref04}. The combination of components, sensors and actuators connected over the IoT enables applications to be created to handle different tasks.

A CPS can have different functionalities, where several metrics (e.g. utilization of hardware components) can be used to measure, resulting in changing the physical environment. An example could be measuring the distance between two objects.
A proximity sensor might measure the distance between a person and an automated teller machine (ATM), based on which, if it is less than 0.5 meters, the withdrawal is allowed. Since such a proximity sensor is placed outdoors, it cannot be connected to the power grid in all cases, thus a power source must be found elsewhere. This can be achieved by using an accumulator or battery. As illustrated in this example, CPS may include devices that are constrained by limited energy resources. Hence, it can be pointed out that a CPS must be functional on the one hand to fulfil a certain task, while on the other hand it has to perform well to ensure the longest possible lifetime. To enable a CPS to complete a task, various  approaches exist on how to design the control loops required. Each of those design approaches is specialized in certain functionalities and comes along with advantages and disadvantages. However, it is important to note that design approaches of control loops are only one responsible part of power consumption, as other factors such as hardware selection, security or communication protocols used also contribute. Therefore, we have addressed the issue of how power consumption can be reduced and subsequently optimized by choosing the most suitable design approach for control loops. 

In order to answer the question regarding the relationship between control loop design approaches and energy consumption, a concept is preparing. To the best of our knowledge, we found related work (see Section II) where design approaches are compared, but none in context with power consumption associated with a comparison of those approaches related to control loops. In this paper we present an experimental setup where we measure the power consumption of relevant hardware components within a CPS, implementing two different design approaches (time-driven and event-driven). This CPS consists of a sensor, an actuator and an IoT framework. The purpose of the IoT framework is to control the way of how new components enter the CPS. Furthermore, it manages the interaction between already participating components to measure and change the physical environment. In this case the Arrowhead Framework is used as representational IoT framework, as it is an open source project written in Java, which can be extended with new functionalities to support more use cases \cite{Delsing2017}. After the implementation of the prototype, measurements were taken to see if a difference in the power consumption exists between the different approaches for control loops. The motivation is to determine whether power consumption savings can be achieved by specifically implementing design approaches of control loops in CPS. 

This paper builds on the work of Sailer \cite{Sailer2019} where the trade-off between security and power consumption for a CPS was investigated. Therefore, in this paper the experimental setup is extended to enable the comparison of two design approaches, while using a representative IoT framework. The aim of this paper is to measure the power consumption of different design approaches for implementing control loops in a simple use case. The main focus is on the design approaches, therefore other factors, such as the selection of protocols or security are not taken into account in these initial measurements, but are planned for future work. In summary, the contribution of this paper can be structured as follows: 
\begin{itemize}
	\item identification and analyses of two representative design approaches
	\item development of a prototype using an IoT framework 
	\item preliminary measurement of power consumption based on identified design approaches and developed prototypes
\end{itemize}

The remainder of this paper is organized as follows: Section II summarizes the related work in the field and points out the background of this paper. Next, in Section III we present an overview and a comparison of two approaches, based on an initial literature review. Furthermore, it explains the implementation of a prototype, followed by the presentation of preliminary measurements and the discussion of those results. Finally, Section IV gives an outline of future work in the field and concludes this paper.

\section{Related Work}

Many studies about designing a CPS have been conducted. 
Gacrilescu et al. \cite{gavrilescu} carried out research on a framework representing an event-driven simulator to configure Programmable System on Chip (PSoC) within a CPS. A classification of a CPS in design (architecture, modelling, tools, simulator, verification), aspects and issues (security, resiliency, reliability, Quality of Services, real-time requirements), applications (such as Vehicular systems, Medical systems, Smart homes, Scheduling, Social network and gaming, Power Grid etc) and challenges has been presented by Khaitan and McCalley \cite{khaitan2014design}. 
The research of Mo et al. \cite{mo} focus on mobile actuators by introducing a new event-driven method, which provides a required level of control accuracy, simultaneously reducing the energy consumption of the actuators, with restricted action delay is guaranteed. However, up to this time research has mainly been conducted in the field of stationary actuators, for example with focus to the actuator control problem, in which the requirements of control accuracy should be fulfilled \cite{yeh2010autonomous}. If this is taken into account, the control quality can be further increased by reducing the action delays \cite{mo2018distributed} or the packet loss rate \cite{cao2009building}. Furthermore, inaccurate system parameters should be identified to prevent them \cite{chen2010distributed}.

Further there are studies comparing event-driven and time (scheduled)-driven approaches for CPS. While Albert et al. \cite{albert2004comparison} compare these and propose hybrid systems as solution, Dai et al. \cite{dai2015time} concentrate on time-stamped event execution within industrial CPS for the IEC 61499 standard. 

Regarding using an IoT framework a huge number of studies exists. Commercially available ones are summarized by Derhamy et al. \cite{Derhamy2015}, including Arrowhead \cite{Delsing2017}, AllJoyn \cite{Alliance2016} or IoTivity \cite{Linux2019}. Ammar et al. \cite{Ammar2018} analysed among other the IoT frameworks SmartThings Samsung \cite{Smart2019}, AWS IoT Amazon \cite{Amazon2019} or Azure IoT Microsoft \cite{Microsoft2019} focusing on different properties like architecture components, hardware and software dependencies, access control or communication.

Additionally, there are many studies looking at power measurement in a CPS considering aspects like security, use of Wireless Local Area Network (WLAN) vs. Local Area Network (LAN) or different protocols.
Carrara et al. \cite{Carrara2010} focus on implementing an IoT-based management program to collect temperature and humidity data, while Tauber et al. \cite{Tauber2011} investigate energy efficiency and performance in a WLAN to identify upper and lower bounds of energy efficiency due to different data flow characteristics. The influence of different security settings on the power consumption was investigated by Sailer et al. \cite{Sailer2019} by showing the impact of strength of security in regard to the power needed. An application model with the possibility to describe the optimal schedule using mathematical formulas to achieve maximum energy savings was presented by Jiang et al. \cite{jiang2008energy}. Shih et al. \cite{shih2002wake} present a strategy for achieving power savings in battery-powered devices and compared it with other strategy approaches. Further research on energy consumption were carried out by Tauber et al. \cite{Tauber2012, Tauber2012.1} and Kansal et al. \cite{Kansal2010}. Furthermore, there are many papers dealing with the issue related to event vs. time-driven energy performance, but their focus is mainly on hardware issues, monitoring frameworks \cite{dillon2011web, makedon} or are looking at a specific scenario or application areas (e.g. smart grids) \cite{cecati}. In our extensive research we did not come across research that is directly related to the proposed scheme. Therefore, with our concept we would like to extend the existing studies of comparisons of design approaches in control loops through establishing a connection to the topic of power consumption. For this purpose, the setup from previous work \cite{Sailer2019} will be extended by control loops, which will be implemented using existing design approaches. The result should be a ranking of the power consumption of the different approaches of control loops in the context of different use cases. 

\section{Methodology approach}

In this section we present the experimental setup for the power consumption measurements within a CPS. 
To test our concept, a CPS is built with the minimum requirements, consisting of a single sensor, actuator and controller (IoT framework). This means that one participant in this CPS measures the physical environment, another changes it, while the third is responsible for controlling the entire interaction between the sensor and actuator. This setup represents the smallest setup of interacting components that are needed to form a CPS, measure and change the physical world. After careful consideration, it was decided that this CPS was sufficient to show whether a difference is made by selecting a different design approach. It is also possible to extend the CPS with additional components to gain complexity, ensuring scalability. A measurement would be required for each additional component, but the measurement method would not change, regardless of the number of components within a CPS. In summary, we have chosen this simple CPS to proof the feasibility of the concept in the smallest possible CPS. 

\begin{figure*}[!h]
	\centerline{\includegraphics[width=0.85\textwidth]{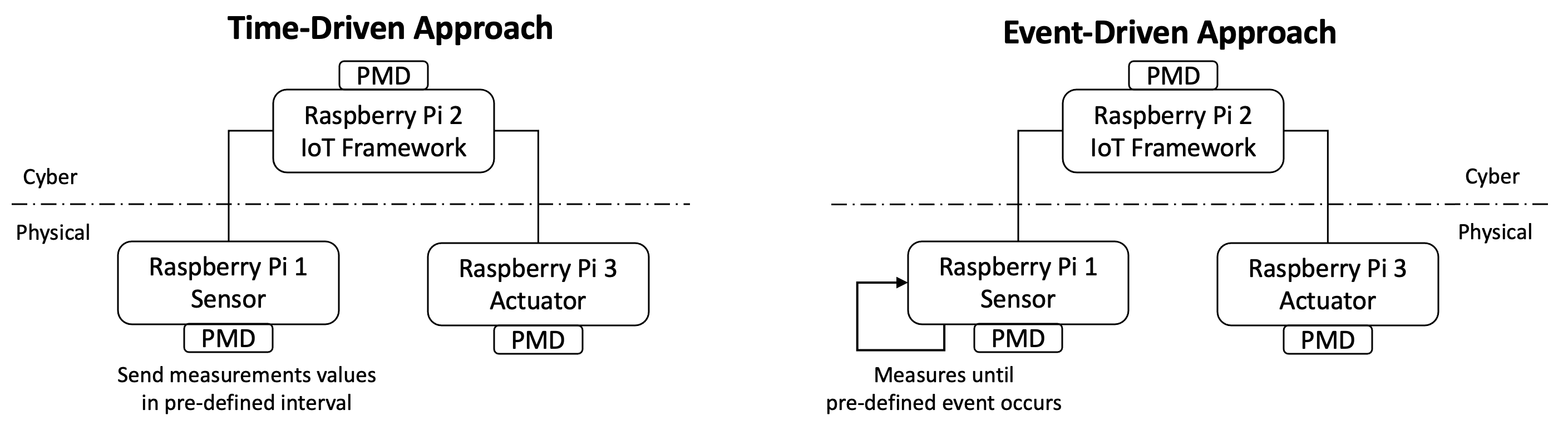}}
	\caption{Prototype architecture consists of Raspberry Pis and power measurement devices (PMD) to compare the time-driven and event-driven design approach. }
	\label{timeevent}
\end{figure*}

\subsection{Use Case}

In principle, a CPS consists of several components interacting with each other, measuring and changing the environment conditions (if necessary). The minimum possible implementation of this definition would be to build a CPS with one component including a sensor, one component including an actuator and a IoT framework. The IoT framework manages the communication and interaction of the sensor and the actuator. For this purpose, authorisation rules are specified to avoid connections between invalid components. To show any difference in the power consumption between design approaches of control loops, we have implemented a use case with minimal requirements using two different design approaches. Furthermore, Table \ref{usecasestable} provides an overview of possible actions to be performed in a CPS. For this purpose, a sensor and an actuator were identified, followed by the functionality they should perform. 

These use cases are scenarios of real-world situations. Since it is difficult to obtain perfect measurement conditions in the real world, these use cases are simulated. This means the sensor sends pre-defined data to the actuator, which simulates an action, if required. Therefore, it is possible to repeat and adjust the measurements to make the results comparable. 

However, if sensors and actuators were actually used for the measurements to compare the results, a laboratory experiment would have to be carried out. A laboratory experiment offers the possibility to specify perfect conditions for measurements and to change them if necessary. To explain this with an example, Use Case 1 from Table \ref{usecasestable} is taken. In a real environment, it would not be possible to influence the measurement values and the actions of the actuator to keep exactly the same schedule twice. Measuring the temperature in a room for one hour, it is probable to have different temperature values in the following hour. Even if the measurements are taken at the same time, the conditions are not identical because the experimental setups cannot be set up in the same location. It is highly possible that an experiment set up is placed closer to the air-conditioning system. This was only one example of the difficulty of comparing measurements in the real world. The purpose of this paper is to show if there is an impact on power consumption even in a minimal CPS by comparing only two design approaches. These results can later be extended by a range of factors, such as WLAN instead of LAN, other protocols or different security settings, to generate an insight into how these could be used in more complex environments. 

\begin{table}[h]
	\caption{Overview of representative use cases}
	\label{usecasestable}
	\begin{tabular}{|l|l|l|l|}
		\hline
		& \textbf{Actuator}                                         & \textbf{Sensor}                                                          & \textbf{Action}                                                                                                          \\ \hline
		\textbf{\begin{tabular}[c]{@{}l@{}}Use \\ Case 1\end{tabular}} & Air Condition                                             & \begin{tabular}[c]{@{}l@{}}Temperature \\ Sensor Inside\end{tabular}     & \begin{tabular}[c]{@{}l@{}}Switch the air conditioning \\ system on/off by reaching \\ certain temperatures\end{tabular} \\ \hline
		\textbf{\begin{tabular}[c]{@{}l@{}}Use\\ Case 2\end{tabular}}  & \begin{tabular}[c]{@{}l@{}}Heating \\ System\end{tabular} & \begin{tabular}[c]{@{}l@{}}Temperature \\ Sensor Outside\end{tabular} & \begin{tabular}[c]{@{}l@{}}Switch the heating system \\ on/off by reaching certain \\ temperatures outside\end{tabular}  \\ \hline
		\textbf{\begin{tabular}[c]{@{}l@{}}Use\\ Case 3\end{tabular}}  & \begin{tabular}[c]{@{}l@{}}Fan/\\ Humidifier\end{tabular} & \begin{tabular}[c]{@{}l@{}}Humiditiy \\ Sensor\end{tabular}              & \begin{tabular}[c]{@{}l@{}}Switch the fan/humidifier \\ on/off by reaching certain \\ limits\end{tabular}                \\ \hline
		\textbf{\begin{tabular}[c]{@{}l@{}}Use\\ Case 4\end{tabular}}  & Camera                                                    & \begin{tabular}[c]{@{}l@{}}Motion  Sensor\end{tabular}                 & \begin{tabular}[c]{@{}l@{}}Video surveillance of e.g. \\ pets at home\end{tabular}                                       \\ \hline
	\end{tabular}
\end{table}

\subsection{Design Approaches for Control Loop}

In an initial literature review we identified two different design approaches, time-driven and event-driven, which offer different procedures since tasks are processed using various strategies. Managing a CPS requires the use of design approaches for control loops to ensure the CPS is doing exactly what it is supposed to do. This means the process for handling tasks is predefined in terms of when a component should do what. In order to manage such a system, appropriate rules for these interactions need to be designed and established. The challenge in designing control loops of CPS is to specify the expected behaviour of the computing components and their impact on the physical environment. Therefore, the programming language has to be considered, since it has to provide an integration of time-driven computation and event-driven computation to enable an asynchronous dynamic [27]. 

In Fig. \ref{timeevent} two design approaches for control loops are compared. The left one is the time-driven approach and the right one is the event-driven approach. In the following, we will explain the two selected approaches in more detail. 

The time-driven approach focuses on a time frame and the execution of different tasks at certain predefined times, which are repeating. If this approach is considered with respect to a real-time system, it can be claimed such a system has a deterministic behaviour. In other words, it can be determined in advance how this system has to behave. As a result, it is possible to plan how the system or subsystem will be executed, since this takes place periodically. When planning, it is important the engineer is aware of that tasks that share a resource cannot be executed simultaneously. Another important issue is to distinguish whether a task is time critical (hard time condition) or not (soft time condition). Fig. \ref{timeevent} shows the process of time-driven approach, in which the actuator requests the values of the sensor in a predefined time interval to detect if an action is needed. This implies each measured value is sent to the actuator via the IoT framework. To make such a system work, the following four requirements must be fulfilled: 
\begin{itemize}
	\item \textbf{Timeliness} - correctness of the system is assumed by implementing the logic for hard time conditions, since result and time of the result creation are important
	\item \textbf{Reliability} - It is essential to ensure there is an emergency solution in case of failure to prevent loss of life, irreparable damage or material damage. 
	\item \textbf{Availability} - It is expected the system will be operational at the specified time. 
	\item  \textbf{Predictability} - All actions to be performed by time interval must be known beforehand. This is necessary to ensure trouble-free action scheduling by taking all necessary precautions. 
	
\end{itemize}

\begin{figure}[h]
	\centerline{\includegraphics[height=0.085\textheight]{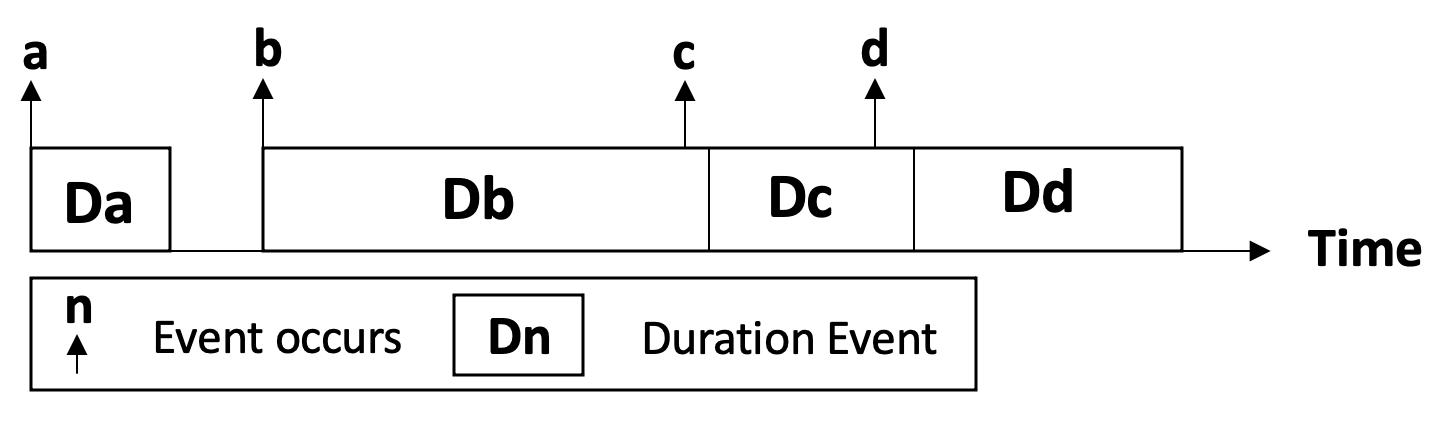}}
	\caption{Timeline indicator when processing various events, showing the time of occurrence of an event, the duration and the end time.}
	\label{timeline}
\end{figure}

In contrast, with the event-driven approach, no predefined interval exists for sending the measurement data, as this is only transmitted from the sensor to the actuator when an event occurs to activate the system. One characteristic of this approach is the possibility that multiple events can occur at the same time. In such a scenario the next upcoming event is chosen for execution. Often event queues are used for this purpose, which, if configured incorrectly, represent a high risk. A further problem of event-driven systems can be the time delay when several events are processed, which is shown in Fig. \ref{timeline}. As a result, such systems can exhibit a jitter, which means that a desired signal is delayed in time causing the process to be delayed.

This comparison shows how important it is to identify the purpose of a CPS before creating it. Not every approach is suitable for each use case. Since both design approaches for control loops can be possible solutions for the implementation of use cases, it should be investigated which of them consumes less power. Especially in such a situation, power consumption could be a criterion for deciding which approach to choose. 

\subsection{Experimental Setup}

In this section we present the experimental setup for the power consumption measurements. Although choice of hardware can already have an influence on power consumption, we selected same devices, equipped with identical operating system, software and services. Therefore, to test our concept, we chose three  Raspberry Pis 3 Model B+, which are used to deploy a CPS consisting of a sensor, an actuator and the IoT framework Arrowhead. As operation system we chose Rasbian Stretch Lite to reduce unnecessary factors like, e.g. resources used for running Graphical User Interface (GUI). To avoid network latency (or any unnecessary external influences) the devices were connected via a switch in a LAN environment. Each Raspberry Pi is equipped with a power measurement device (PMD) to determine the energy consumption. We decided to use the PiLogger One \cite{Pilogger} as PMD, which can be attached directly to a Raspberry Pi and will be connected via Inter-Integrated Circuit (I$^{2}$C) bus. An advantage of the PiLogger One is the possibility to use time precise intervals for the measurements, as it offers its own time base. The measured values are stored and can be evaluated afterwards. Based on Fig. \ref{timeevent} it is visible which Raspberry Pi assumes which functionality in the experimental setup. 

As shown in Fig. \ref{timeevent} the sensor and the actuator are connected over an IoT framework. 
The Arrowhead Framework \cite{Delsing2017} is used as a representative IoT framework to facilitate the creation of local automation clouds, which includes devices, various application-specific systems and services to perform the automation tasks and provide a boundary to the open internet and the outside activities. This way Arrowhead is enabling local (on-site or private), real time performance and security, paired with simple and cheap engineering, while enabling scalability through multi-cloud interactions. Further it provides an architecture composed of three mandatory core systems: Service Registry System, Orchestration System and Authorisation System.

In this use case the components sensor and actuator are forming a CPS by using the Arrowhead Framework to become part of it and controlling the interactions. Once the on-boarding procedure has been completed successfully and the two IoT-devices are part of the CPS their main purpose is to measure the temperature of a physical room and to regulate its temperature (if necessary). 
The experimental setup emulates a network in which data is collected with a sensor and is sent via an Ethernet switch to the Arrowhead local cloud by using the the Hypertext Transfer Protocol (HTTP) protocol. The components are explained in detail below:
\begin{itemize}
	\item Actuator simulates an air-conditioning system to cool down a physical room if necessary. 
	\item Sensor measures temperature of a physical environment. 
	\item Service Registry is responsible to register new services. 
	\item Authorisation System is responsible for authorisation rules, which regulate, which components are allowed to interact with each other within the CPS. Every time a component sends a request to the Arrowhead Framework, it verifies if this component is allowed to do so.
	\item Orchestration System is responsible for handling any request by any component. Every time a component needs information about the CPS, the orchestration handles the request and make sure that the right information is passed to the requesting component.
\end{itemize}

It is necessary to deploy the Arrowhead Core Systems in a local cloud environment and register the sensor and actuator components before this CPS can work. After these steps have been successfully performed, it is possible to execute the CPS using each of the presented design approaches. The time-driven approach uses a sensor to measure the temperature at defined intervals, e.g. every ten minutes and send this value to the Arrowhead Framework. Within the framework, the value is first passed to the Orchestration System, which requests the Service Registry to verify whether there is a suitable actuator. After successful response, Authorisation System is used to check whether the sensor is allowed to communicate with the matching actuator. If this is confirmed, the value is forwarded to the actuator, in this case the air condition, who subsequently decides whether an action has to be set or not. This means that this process is carried out continuously, irrespective of the measured temperature value. 
On the contrary to this, the event-driven approach offers the possibility of a prior check whether the measured temperature value has been reached or exceeded a predefined limit. If the value is below this limit, no action is required and the measurement is repeated. This action is continued until the value is above the limit, which means the actuator has to be informed about performing an appropriate action.  In this case, the value is sent to the Arrowhead Framework and the same procedure is performed as in the time-driven approach.

To reinforce our presented concept, we have made preliminary measurements including the two presented approaches. Fig. \ref{timeevent} shows the experimental setup for both implementations including the comparison of the approaches. A total of 22 test runs with 100 measurements each were carried out, whereas the measured values were predefined in order to be able to reproduce each test run under the same conditions. In each of the 100 measurements per test run, a predefined value was read from an array containing 50 values below and 50 above a predefined limit, which was set at the value 25. Therefore, measurements 1 to 50 do not activate the actuator, in contrast to measurements 51 to 100. 
The first test run was a warm-up and was not taken into account for the exploitation. Furthermore, all unnecessary services on the Raspberry Pis were turned off to avoid unnecessary power consumption.

\subsection{Evaluation}
\begin{figure}[!h] 
	\centering 
	\includegraphics[width=0.5\textwidth]{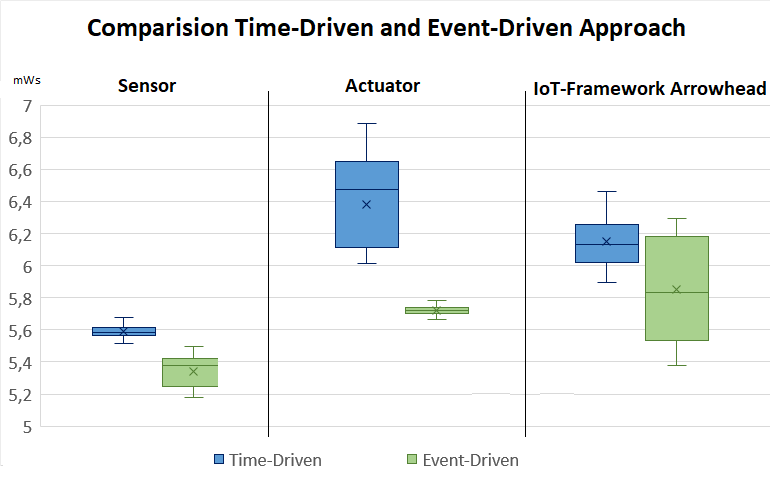} 
	\caption{Overview of the power consumption in Milliwattseconds (mWs) of the testruns performed. To compare the time-driven and event-driven approach the components Sensor, Actuator and Arrowhead Framework were measured.}
	\label{comparision}
\end{figure}

The results of the measurements are summarized in Fig. \ref{comparision}, whereby the time-driven approach is shown in blue and the event-driven approach in green. The results show a significant difference between the design approaches for the devices representing the sensor and actuator. This is because the sensor in the time-driven approach establishes a connection to the IoT framework for each measurement in order to search for the actuator. In the event-driven approach this connection was only established when the specified value achieved a defined limit. The test values were chosen regularly, whereby the first half was below the limit and the second half was above it. Further observation of the test runs showed that a test run in the time-driven approach needed an average of 125 seconds to take 100 measurements, while the event-driven approach needed only 114 seconds, showing a time savings in addition to the power savings. 

In contrast, the measurements of the device with the Arrowhead Framework gave a similar result, observing the time-driven approach needs more power.
However, it also shows that in the event driven approach the framework has a high dispersion, which implies that parts of the energy consumption of both approaches overlap. At this stage, we have no clear indication of this behaviour. Moreover, it was noticed in all cases that the initial test runs consumed more power, which can be interpreted as a result of the warm-up phase. Therefore, a longer warm-up phase is planned for further measurements. Summarising, these first measurements show that choosing a specific design approach has direct impact on the resulting power consumption. Our results show that the event-driven approach consumed approximately 7\% less power, compared to the time-driven approach.

\section{Conclusion and Future Work}

In this paper, we show the impact of the selection of the design approach of control loops in a CPS on power consumption. Therefore, we identified two design approaches  in an initial literature review, which were explained in detail. In addition to that, we describe the minimum requirements to build and measure the power consumption of design approaches of control loops of a CPS. Hence, we performed power consumption measurements with the developed prototype using Raspberry Pi 3 Model B+ as devices, each equipped with a PiLogger One as PMD. Furthermore, we presented initial measurements of the power consumption by using the time-driven and event-driven approach to design control loops. Finally, our experimental study indicated that the event-driven approach consumes 7\% less power to process the same task comparing with the time-driven approach.

The main contribution of this paper is the initial measurements of the impact of power consumption on the selection of design approaches for control loops. A difference in power consumption by considering different design approaches has been shown. This will be enhanced in future work by addressing further existing design approaches and implementing other use cases to make additional measurements. Since the results of the initial measurements of the IoT framework were partly overlapping, an aspect of swinging in control loops will be taken into account for future measurements, as well as the extension of the warm-up phase. Furthermore, other IoT devices, such as Arduino, can be used in order to examine the connection to power consumption from the hardware side as well. In addition, a more detailed investigation by extending the protocols, security and algorithms will be carried out in continuing work. Furthermore, a model for power measurements should be developed in order to be generalized for further research and to be used in industry.

In addition, we are considering how to conceptually use the different design approaches of control loops to create a dynamic and autonomic system. Such a system will be able to identify the current situation regarding power consumption and switch to the most suitable design approach for control loops. For this purpose, all impacts on power consumption must first be identified to enable the control loops to be adjusted accordingly. To create a self-adaptable system, tools such as the interactive framework Monitor-Analysis-Plan-Execute over a shared Knowledge (MAPE-K) \cite{nguyen2015self} can be used.

\section*{Acknowledgment}
\vspace{-1mm}
The research has been carried out in the context of the EFRE project MIT 4.0 (FE02), funded by IWB-EFRE 2014-2020 coordinated by Forschung Burgenland GmbH.

\bibliography{conference}
\bibliographystyle{ieeetr}

\end{document}